\documentclass[aps,prd,floats,showpacs,nofootinbib,superscriptaddress,preprint,tightenlines]{revtex4}
\usepackage[centertags]{amsmath}
\usepackage{amsfonts}
\usepackage{amssymb}
\usepackage{amsthm}
\usepackage{newlfont}
\usepackage{hyperref}
\def\dif{\textrm{d}}

\def\bea{\begin{eqnarray}}
\def\eea{\end{eqnarray}}

\usepackage{graphicx}
\def\be{\nopagebreak[3]\begin{equation}}
\def\ee{\end{equation}}
\def\ba{\nopagebreak[3]\begin{eqnarray}}
\def\ea{\end{eqnarray}}

\def\d{{\mathrm{d}}}

\def\H{{\cal H}}

\def\l{\lambda}

\newcommand{\f}{\frac}

\usepackage{enumerate}
\def\f{\frac}

\begin{document}
\preprint{\vbox{\baselineskip=12pt \rightline{IGC-11/5-3}}}

\title{%Consequences of different quantizations
Loop quantum cosmology of  $k$=1 FRW: A tale of two bounces}
\author{Alejandro Corichi}\email{corichi@matmor.unam.mx}
\affiliation{Instituto de Matem\'aticas,
Unidad Morelia, Universidad Nacional Aut\'onoma de
M\'exico, UNAM-Campus Morelia, A. Postal 61-3, Morelia, Michoac\'an 58090,
Mexico}
\affiliation{Center for Fundamental Theory, Institute for Gravitation and the Cosmos,
Pennsylvania State University, University Park
PA 16802, USA}
\author{Asieh Karami}
\email{karami@matmor.unam.mx}
\affiliation{Instituto de F\'{\i}sica y
Matem\'aticas,  Universidad Michoacana de San Nicol\'as de
Hidalgo, Morelia, Michoac\'an, Mexico}
\affiliation{Instituto de Matem\'aticas,
Unidad Morelia, Universidad Nacional Aut\'onoma de M\'exico,
UNAM-Campus Morelia, A. Postal 61-3, Morelia, Michoac\'an 58090,
Mexico}

\begin{abstract}
We consider the $k$=1 Friedman-Robertson-Walker (FRW) model within loop quantum cosmology, paying special attention to the existence of an ambiguity in the quantization process.
In spatially non-flat anisotropic models such as Bianchi II and IX, the standard method of defining the curvature through closed holonomies is not admissible. Instead, one has to implement the quantum constraints by approximating the connection via open holonomies. In the case of flat $k$=0 FRW and Bianchi I models, these two quantization methods coincide, but in the case of the closed $k$=1 FRW model they might yield different quantum theories. In this manuscript we explore these two quantizations and the different effective descriptions they provide of the bouncing cyclic universe. In particular, as we show in detail, the most dramatic difference is that in the theory defined by the new quantization method, there is not one, but {\it two} different bounces through which the cyclic universe alternates. We show that for  a `large' universe, these two bounces are very similar and, therefore, practically indistinguishable, approaching the dynamics of the holonomy based quantum theory.
\end{abstract}

\maketitle

\section{Introduction}

Loop quantum cosmology (LQC) has become in the past years an interesting candidate for a quantum description of the early universe via homogeneous cosmological models \cite{lqc,abl,AA}.
Based on the same quantization methods of loop quantum gravity \cite{lqg}, it has also become a testing ground for different conceptual and technical issues that arise in the full theory. It is perhaps not surprising that the model was first fully understood is the spatially flat $k$=0 FRW cosmological model coupled to the simplest kind of matter, namely a mass-less scalar field that serves as an internal time parameter \cite{aps0,aps1,aps2,polish,slqc,cs:prl}.
It was shown numerically that the big bang singularity is replaced by a quantum bounce \cite{aps2}, that connects an early contraction phase of the universe with the current state of expansion. By means of an exact solvable model, this bounce was then understood to be generic and present for all states of the theory, and the energy density was shown to be absolutely bounded by a critical density $\rho_{\textrm{crit}}$ of the order of the Planck density \cite{slqc}. It was then shown that
semiclassical states after the bounce have to come from states that were also semiclassical well before the bounce \cite{cs:prl,recall,CM}.
This results have also benefited from uniqueness results that warranties the physical consistency of
the theory \cite{cs:unique}.
The same quantization methods were applied to other isotropic models with and without a cosmological constant.
Thus, a closed $k$=1 was extensively studied in \cite{apsv} and \cite{closed}, while the open $k$=-1 was considered in
\cite{open}. A detailed study of singularity resolution for these models was recently completed in \cite{ps:fv}, extending previous results for the flat case \cite{cosmos}. For the flat model,
a cosmological constant was included in \cite{vp} and a massive scalar field in  \cite{massive}, where singularity resolution was also shown to emerge as a feature of the theory.

An extension of this consistent quantization method was successfully implemented for the simples anisotropic cosmology, namely a Bianchi I spacetime in \cite{bianchiI}. It was soon realized that, for anisotropic models with a nontrivial spatial curvature, this quantization method based on considering holonomies along closed loops was no longer applicable. The operator associated to the field strength was no longer well defined on the kinematical Hilbert space of the theory used so far.
The proposal put forward in \cite{bianchiII} was to consider {\it open} holonomies to represent the connection, and
then define the curvature out of the resulting operator. As it turns out, this quantization method has some resemblance to the quantization procedure known as `polymerization' \cite{CVZ-2}. For the quantization of Bianchi IX cosmological models, it was also noted that this `connection quantization' could be successfully implemented \cite{bianchiIX}, and the singularity could also be resolved.

A natural issue that one would like to investigate are the physical consequences of this `new' loop quantization. Do we have the same qualitative behavior as in the holonomy based quantization? This question has been
satisfactorily (but trivially) answered in some cases where both quantizations are available. When the spatial curvature vanishes, as is the case of the $k$=0 FRW and Bianchi I models, both quantization methods coincide \cite{ma,bianchiII} (once one
appropriately fixes a free parameter). It is then quite natural to ask whether the same feature is present in other models where the intrinsic spatial curvature in non-trivial. Is there an important effect that the spatial
curvature carries? In this respect, the $k$=1 FRW model is unique to answer this question since, (to our knowledge) it is the only such model for which both loop quantizations exist.

The purpose of this paper is to explore this issue in detail. More precisely, we shall develop the {\it connection based} quantum theory for a $k$=1 FRW model and explore its more important features by using an effective description of the dynamics. We shall then compare this description with that from the standard --holonomy based-- loop quantization explored in \cite{apsv,closed}, where the effective description has been show to correctly capture the dynamics of semiclassical states \cite{apsv}. Perhaps somewhat surprisingly, what we find is that in the new --connection based-- quantum theory, the corresponding cyclic universe undergoes a series of bounces and recollapses, but now {\it there are two different kind of bounces}. In the cosmic evolution, the universe alternates between these two bounces where both the density and minimum volume differ. Interestingly, for universes that grow to become `large' before the expansion stops,
the two bounces become more similar to each other, so that for a large universe like ours, they become almost indistinguishable.

The structure of this manuscript is the following: In Sec.~\ref{sec:2} we recall the classical $k$=1 model, introducing some new notation. In Sec.~\ref{sec:3} we recall the effective description of the holonomy based quantization and explore some of its consequences. Section \ref{sec:4} is the main section of the paper. In the first part, we develop the loop quantization of the model, and in the second part we consider its effective description. We analyze then some of its consequences. We end in Sec.~\ref{sec:5} with a discussion. In the Appendix we summarize our conventions and the computation of closed holonomies.

\section{Preliminaries: The $k$=1 Cosmology}
\label{sec:2}

The spacetimes under consideration are of the form $M=\Sigma\times \mathbb{R}$, where $\Sigma$ is a topological three-sphere $\mathbb{S}^3$. It is standard to endow $\Sigma$ with a fiducial basis of one-forms
${}^o\!\omega^i_a$ and vectors ${}^o\!e^a_i$. The fiducial metric on $\Sigma$ is then ${}^o\!q_{ab}:=
{}^o\!\omega^i_a\,{}^o\!\omega^j_b\,k_{ij}$, with $k_{ij}$ the Killing-Cartan metric on su(2). Here, the
fiducial metric ${}^o\!q_{ab}$ is the metric of a three sphere of radius $a_0$. The volume of $\Sigma$ with respect to ${}^o\!q_{ab}$ will be denoted by $V_0=2\pi^2\,a_0^3$. We also define the quantity $\ell_0:=V_0^{1/3}$.
It can be written as $\ell_0=:\sigma\, a_0$, where the  quantity $\sigma:=(2\pi^2)^{1/3}$ will appear in many expressions.\footnote{Note that these conventions follow those of \cite{apsv} (compare to \cite{ps:fv}). In spite of this, several of our equations will be
different.}

The isotropic and homogeneous connections and triads can be written in terms of the fiducial quantities as follows,
\be
A_a^i=\f{c}{\ell_0}\,{}^o\!\omega^i_a\qquad ;\qquad E^a_i=\f{p}{\ell^2_0}\sqrt{{}^o\!q}\,{}^o\!e^a_i\, .
\ee
Here, $c$ is dimension-less and $p$ has dimensions of length-squared. The metric and extrinsic curvature can be recovered from the pair $(c,p)$ as follows,
\be
q_{ab}=\f{|p|}{\ell^2_0}\,{}^o\!q_{ab}\qquad \mathrm{and}\qquad \gamma K_{ab}=\left(c-\f{\ell_0}{2}\right)\f{|p|}{\ell^2_0}\,{}^o\!q_{ab}
\ee
Note that the total volume $V$ of the hypersurface $\Sigma$ is given by $V=|p|^{3/2}$. The Poisson bracket for the
phase space variables $(c,p)$ is given, as in the $k$=0 case by,
\be
\{c,p\} = \f{8\pi G\gamma}{3}\, ,
\ee
with $\gamma$ the Barbero-Immirzi parameter.
From here, one can calculate the curvature $F^k_{ab}$ of the connection $A_a^i$ on $\Sigma$ as,
\be
F^k_{ab}=\f{c^2-2\sigma c}{\ell^2_0}\;{\epsilon_{ij}}^k\,{}^o\!\omega^i_a\,{}^o\!\omega^j_b
\ee
The only relevant constraint is the Hamiltonian constraint that has the form,
\be
{\cal H}_{\textrm{grav}}=\int_\Sigma\d^3\!x\,\left[{\epsilon^{ij}}_k\,e^{-1}\,E^a_iE^b_j\,F^k_{ab}-
2(1+\gamma^2)e^{-1}\,E^a_iE^b_j\,K^i_{[a}K^j_{b]}\right]
\ee
where $e=\sqrt{|\textrm{det}E|}$, and $K^i_a$ is the extrinsic curvature. By means of the relation
$A_a^i=\Gamma^i_a+\gamma K_a^i$, with $\Gamma^i_a$ the spin-connection compatible with the triad, we can re-express the second term of the Hamiltonian constraint as,
\be
E^a_iE^b_j\,K^i_{[a}K^j_{b]}=\f{1}{2\gamma^2}{\epsilon^{ij}}_k\,E^{a}_i E^{b}_j(F_{ab}^k-\Omega_{ab}^k)\, .
\ee
Here $\Omega_{ab}^k$ is the curvature of the spin-connection $\Gamma^i_a$.  The advantage of this substitution is that for this model, this expression has a simple form,
\be
\Omega_{ab}^k=-\f{1}{a_0^2}\,{\epsilon_{ij}}^k\, {}^o\!\omega^i_a\,{}^o\!\omega^j_b
\ee
With this, the gravitational constraint can be reduced to,
\be
{\cal H}_{\textrm{grav}}=-\f{3}{8\pi G\gamma^2}\,\sqrt{|p|}\left[(c-\sigma)^2 + \gamma^2\sigma^2\right]
\ee
It is convenient to introduce new variables \cite{slqc}: $\beta:=c/|p|^{1/2}$ and $V=p^{3/2}$. The
quantity $V$ is just the volume of $\Sigma$ and $\beta$ is its canonically conjugate,
\be
\{\beta,V\} = 4\pi G\gamma
\ee
We can then compute the evolution equations of $V$ and $\beta$ in order to find interesting geometrical scalars.
Then,
\be
\dot{V}=\{V,{\cal H}_{\textrm{grav}}\}= \f{3}{\gamma}\left(\beta V - \sigma V^{2/3}\right)
\ee
from which we can find the standard Friedman equation using the constraint equation ${\cal H}=
{\cal H}_{\textrm{grav}} + {\cal H}_{\textrm{matt}}\approx 0$ and ${\cal H}_{\textrm{matt}}=V\rho$,
\be
H^2:=\left(\f{\dot{V}}{3V}\right)^2=\frac{8\pi G}{3} \,\rho-\frac{\sigma^2}{V^{2/3}}\, .
\label{class-frid}
\ee
We can now compute $\dot{\beta}=\{\beta,{\cal H}\}$,
\be
\dot{\beta}:= -\f{3}{2\gamma}\left[\beta^2-\f{4}{3}\sigma\beta V^{-1/3} + \f{1}{3}(1+\gamma^2)\sigma^2 V^{-2/3}\right]
+ 4\pi G\gamma P\label{dot-theta-clas}
\ee
where we have used the standard definition of pressure as $P:=\f{\partial {\cal H}_{\textrm{matt}}}{\partial V}$.
We can readily find the time evolution of the expansion parameter $\theta= 3 H$ as,
\be
\dot{\theta}=4\pi G (\rho- 3P)- \f{3\;\sigma^2}{V^{2/3}}
\ee
From Eq.~(\ref{class-frid}) we can see that the condition for a turnaround point, namely when $H=0$ is that the density satisfies $\rho_{\textrm{turn}}:= \frac{3}{8\pi G}\;\f{\sigma^2}{V^{2/3}}$. This is the point where
the Hubble parameter vanishes.  From (\ref{dot-theta-clas}) we see that, if $P>-\rho/3$ then $\dot\theta<0$ at the turnaround point, which means that there is a transition from an expanding phase (where $\theta>0$) to a contracting phase (where $\theta<0$), so it corresponds to a point of {\it re-collapse}.

\section{Loop quantization I: The holonomy way}
\label{sec:3}

This section has two parts. In the first one, we recall the effective equations for the quantization of the $k$=1 model as developed in Ref.\cite{apsv}, and explore some of its consequences for arbitrary matter content. In the second part we restrict our attention to the case of a mass-less scalar field.
%--------------------------------------

\subsection{Effective equations for holonomy-based quantization}

The basic strategy of loop quantization is that the effects of quantum geometry are manifested by means of holonomies
around closed loops that carry the information about the field strength of the connection. As is shown in detail in the Appendix, the curvature takes then the form,
\be
{}^{\lambda}\!F_{ab}^k=\frac{\sin^2\bar\mu(c-\sigma)-\sin^2(\bar\mu\sigma)}{\bar\mu^2\ell_o^2}
\;{\epsilon_{ij}}^k\,{}^o\!\omega^i_a\,{}^o\!\omega^j_b
\ee
where $\bar\mu=\sqrt{\lambda^2/|p|}$.
In terms of the new variables $\beta=c|p|^{-1/2}$ and $V=|p|^{3/2}$, it can be written as,
\be
{}^{\lambda}\!F_{ab}^k = \f{V^{2/3}}{\lambda^2\ell^2_0}\left[\sin^2(\lambda\beta - D) - \sin^2D\right]
\;{\epsilon_{ij}}^k\,{}^o\!\omega^i_a\,{}^o\!\omega^j_b
\ee
where we have defined $D:=\lambda\sigma/V^{1/3}$. With this form of the curvature as defined by closed holonomies, and
neglecting the so called inverse triad corrections, one can arrive at the form of the effective Hamiltonian,
\be
\mathcal{H}_{\textrm{eff}}=-\frac{3}{8\pi G\gamma^2\lambda^2}V\left[\sin^2(\lambda\beta-D)-\sin^2D+(1+\gamma^2)D^2\right]+\rho V
\ee
We can now compute the equations of motion from the effective Hamiltonian as,
$$\dot{V}=\{V,\mathcal{H}_{\textrm{eff}}\}=\{V,\beta\}\frac{\partial\mathcal{H}_{\textrm{eff}}}{\partial\beta}=\frac{3}{\lambda\gamma}V\sin(\lambda\beta-D)\cos(\lambda\beta-D)\, .
$$
From here, we can find the expansion as,
\be
\theta=\frac{\dot{V}}{V}=\frac{3}{\lambda\gamma}\sin(\lambda\beta-D)\cos(\lambda\beta-D)=\frac{3}{2\lambda\gamma}\sin2(\lambda\beta-D)\label{exp-1}\, .
\ee
From the above equation we can see that the absolute value of expansion has an absolute upper limit equal to $|\theta|\leq 3/2\lambda\gamma$. We can now compute the modified, {\it effective Friedman equation}, by computing $H^2=\frac{\theta^2}{9}$,
\be
\begin{split}
H^2 & = \frac{1}{\lambda^2\gamma^2}\left(\frac{8\pi G\gamma^2\lambda^2}{3}\rho+\sin^2D-(1+\gamma^2)D^2\right)
\left(1-\frac{8\pi G\gamma^2\lambda^2}{3}\rho-\sin^2D+(1+\gamma^2)D^2\right)\\
&=\frac{8\pi G}{3}(\rho-\rho_1)\left(1-\frac{\rho-\rho_1}{\rho_{crit}}\right)
\end{split}\label{eff-frid-1}
\ee
where
$\rho_1=\rho_{\textrm{crit}}[(1+\gamma^2)D^2-\sin^2D]$ and
$\rho_{\textrm{crit}}=3/(8\pi G\gamma^2\lambda^2)$ is the  {\it critical density} of the $k=0$ FRW model.
 We can immediately note from Eq.~(\ref{eff-frid-1}) that there are two points where the Hubble
parameter $H$ vanishes and the Universe has a turnaround. The first one corresponds to the point $\rho=\rho_1$. %represents a lower bound for density, and it signals the classical turnaround where the Universe stops expanding and begins to re-collapse.
Note that $\rho_1$, in the limit $\lambda \to 0$, tends to $\rho_1\mapsto \f{3}{8\pi G}\,\f{\sigma^2}{V^{2/3}}$, which is the classical value for re-collapse as given by Eq.~(\ref{class-frid}). Thus, in the limit of large volumes
one expects $\rho_1$ to represent the density at re-collapse.
The second value for density where the Hubble parameter vanishes is given by $\rho=\rho_{\textrm{crit}}+
\rho_1$.
%It represents an upper value for density and corresponds to the point where the Universe stops
%its collapse and starts to expand again. The quantum bounce.
Note that these densities, where there is a turnaround, is not an
universal constant for all trajectories as was the case for the $k$=0 model (for the bounce at $\rho=\rho_{\textrm{crit}}$). Instead,
the quantity $\rho_1$ is a function of volume and depends on each individual trajectory. The second density for
turnaround is
bounded below by $\rho_{\textrm{crit}}$.\footnote{Also note that since $\rho_1$ depends explicitly on the volume, the values it takes at the bounce and classical turnaround point are different, so it could happen that $\rho=\rho_1$ is actually larger than in the other root, and it corresponds to the bounce while $\rho=\rho_{\textrm{crit}}+\rho_1$ corresponds to a re-collapse \cite{CKM}.}
There is an alternate way of analyzing  the two turnaround points. From the expression of the expansion (\ref{exp-1}) we can see that the Hubble parameter vanishes when
\be
\sin2(\lambda\beta - D)=\sin(\lambda\beta -D)\cos(\lambda\beta - D)= 0
\ee
There are two possibilities for this.

i) When
$
\lambda\beta -D = \f{(2n+1)}{2}\,\pi\, ,
$\\
for $n$ integer, which corresponds to $\rho=\rho_{\textrm{crit}}+\rho_1$.
The other possibility is,

ii)
$
\lambda\beta - D= m\,\pi
$\\
where $m$ is an integer number.  This corresponds to $\rho=\rho_1$.

In fact,
these considerations suggest that we could define a new variable $\tilde\beta:=\beta - D/\lambda=(c - \sigma)/\sqrt{p}$, that would also be `conjugate' to $V$ ($\{\tilde\beta,V\} =4\pi G\gamma$). In terms of $\tilde\beta$ many expressions would simplify, and it would reduce to $\beta$ in the $k$=0 case.

In order to determine which of the turnaround points corresponds to a bounce and which one to a re-collapse, we need
to consider the rest of the effective equations of motion,
%
%\be
%\dot V=\frac{3}{2\lambda\gamma}V\sin2(\lambda\beta-D)\qquad ;\qquad\dot\phi=\frac{p_{\phi}}{V}\qquad;\qquad
%\dot{p}_\phi=-V\,U_{,\phi}
%\ee
%and,
\be
\begin{split}
\dot\beta&=4\pi G\gamma P\\
%\frac{2\pi G\gamma p_\phi^2}{V^2}+4\pi G\gamma U(\phi)\\
&-\frac{1}{2\gamma\lambda^2}\left[3\sin^2(\lambda\beta-D)-3\sin^2D+D\sin2(\lambda\beta-D)+D\sin2D+(1+\gamma^2)D^2\right]\\
%&=-\frac{4\pi G\gamma p_\phi^2}{V^2}-\frac{D}{2\gamma\lambda^2}\left[\sin2(\lambda\beta-D)+\sin %2D-2(1+\gamma^2)D\right]\\
&=-4\pi G\gamma\left[\rho-\rho_2+P\right]
\end{split}
\ee
where
\be
\rho_2=\frac{\rho_{crit}D}{3}\left[2(1+\gamma^2)D-\sin2(\lambda\beta-D)-\sin 2D\right]
\ee
%
%From the continuity equation we have $p_\phi^2/V^2=\rho+P$ where $P$ is the pressure of the matter.
The Ricci scalar is given by,
\be
\begin{split}
R&=2\dot\theta+\frac{4\theta^2}{3}+\frac{6\sigma^2}{V^{2/3}}\\
&=8\pi G\rho\left(1+2\frac{\rho-\rho_1}{\rho_{\textrm{crit}}}\right)+32\pi G\rho_1\left(1-\frac{\rho-\rho_1}{\rho_{\textrm{crit}}}\right)-24\pi G(P-\rho_3)\left(1-2\frac{\rho-\rho_1}{\rho_{\textrm{crit}}}\right)+\frac{6\sigma^2}{V^{2/3}}
\end{split}
\ee
The time derivative of the expansion is given by,
\be
\dot\theta=\cos2(\lambda\beta-D)\left(\frac{3}{\gamma}\dot\beta+\frac{\theta D}{\gamma\lambda}\right)=\left(\frac{3}{\gamma}\dot\beta+\frac{\theta D}{\gamma\lambda}\right)\left[1-2\frac{\rho-\rho_1}{\rho_{\textrm{crit}}}\right]=-12\pi G\left(\rho-\rho_3+P\right)\left[1-2\frac{\rho-\rho_1}{\rho_{\textrm{crit}}}\right]\label{dot-theta1}
\ee
with
$$\rho_3=\rho_2+\frac{\rho_{crit}D}{3}\sin2(\lambda\beta-D)=\frac{\rho_{crit}D}{3}\left[2(1+\gamma^2)D-\sin 2D\right]$$
Finally, the contracted Ricci curvature appearing in Raychaudhuri equation is given by,
$$
R_{ab}\xi^a\xi^b=-\dot\theta-\frac{1}{3}\theta^2=4\pi G\rho\left(1-4\frac{\rho-\rho_1}{\rho_{crit}}\right)+8\pi G\rho_1\left(1-\frac{\rho-\rho_1}{\rho_{crit}}\right)+12\pi G(P-\rho_3)\left(1-2\frac{\rho-\rho_1}{\rho_{crit}}\right)
$$
It is straightforward to show that the continuity equation $\dot{\rho}+3H(\rho + P)=0$ is also satisfied in this case
\cite{ps:fv}.

Let us now determine the nature of the turnaround points.
From Eq.~(\ref{dot-theta1}) we can see that in case i) above, where
$\theta=0$ and $\rho=\rho_{\textrm{crit}} +\rho_1$, we have then, % Thus, in the i) case
\be
\dot\theta=-\f{1}{\gamma}\dot\beta
\ee
Therefore, the nature of the turnaround is determined by the sign of $\dot\beta$. If $\dot\beta<0$ then
$\dot\theta>0$ and the point corresponds to a {\it bounce}. However, if  $\dot\beta > 0$ then
$\dot\theta < 0$ and the point corresponds to a {\it re-collapse}.

For case ii), again from Eq.~(\ref{dot-theta1}), and using $\theta=0$ and $\rho=\rho_1$ we can see that,
\be
\dot\theta=\f{1}{\gamma}\dot\beta
\ee
Therefore, if $\dot\beta<0$ then
$\dot\theta < 0$ and the point corresponds to a {\it re-collapse}. In the other case, when  $\dot\beta > 0$ then
$\dot\theta < 0$ and the point corresponds to a {\it bounce}.
From this discussion, we can see that the nature of the turnaround points can change if, during the dynamical evolution, $\dot\beta$ changes sign.
This phenomena has indeed been observed in certain cases \cite{CKM}.

\subsection{Concrete example: A massless scalar}

Up until now, we have considered arbitrary matter sources. Let us now restrict our attention to the simplest case of a massless scalar field $\phi$, where the density is given by $\rho=\dot\phi^2/2$ \cite{apsv}.  In this case,
$\dot\beta < 0$ and does not change during the dynamical evolution. This means that the case i) above corresponds to the bounce and case ii) to the re-collapse. In order to
find the minimum and maximum volume we can put the maximum or minimum density in one side of the expression of density to have,
\be
\frac{p_{\phi}^2}{2V_{max}^2}=\rho_{\textrm{crit}}\left[(1+\gamma^2)\frac{\lambda^2\sigma^2}{V^{2/3}_{\textrm{max}}}-\sin^2\frac{\lambda\sigma}{V^{1/3}_{max}}\right]
\label{V_max11}
\ee
and
\be
\frac{p_{\phi}^2}{2V_{\textrm{min}}^2}=\rho_{\textrm{crit}}\left[1+(1+\gamma^2)\frac{\lambda^2\sigma^2}{V^{2/3}_{\textrm{min}}}-\sin^2\frac{\lambda\sigma}{V^{1/3}_{\textrm{min}}}\right]
\label{V_min11}
\ee
From numerical simulations performed in Ref.~\cite{apsv} and analytical considerations for the $k$=0 model \cite{CM},
we know that the constant of the motion $p_\phi$ determines how {\it semiclassical} the state is. To be precise,
as one increases the value of $p_\phi$, in natural Planck units, it becomes easier to construct semiclassical states
peaked on that value of the field momenta. It is then natural to expect that $p_\phi$ measures in a way, how {\it large} the Universe can grow before the re-collapse phase starts. That is certainly true for the classical equations
of motion. Since we expect that the classical equations are a good approximation to the effective equations of motion in the low density regime, the volume at with the expansion stops should coincide when this transition happens at low densities in Planck units. Therefore, let us assume that $V_{\textrm{max}}^{1/3}\gg \sigma\lambda$, which means,
\be
p^2_\phi = 2V_{\textrm{max}}^{2}\,\rho_{\textrm{crit}}\left[(1+\gamma^2)\,\f{\lambda^2\sigma^2}{V_{\textrm{max}}^{2/3}}-
\sin^2\left(
\f{\lambda\sigma}{V_{\textrm{max}}^{1/3}}\right)\right]\approx 2 V_{\textrm{max}}^{2}\,\rho_{\textrm{crit}} \f{\gamma^2\lambda^2\sigma^2}{V^{2/3}_{\textrm{max}}}
\ee
from which we can see that the maximum value of volume approaches the classical value
\be
V_{\textrm{max}}=\left( \f{64\pi G}{3\sigma^2}\right)^{3/4}\,p_\phi^{3/2}
\ee
from above. Let us now estimate the value of the bounce in the same regime, where the value of $p_\phi$ is large.
\be
p^2_\phi = 2V_{\textrm{min}}^{2}\,
\rho_{\textrm{crit}}\left[1+(1+\gamma^2)\,\f{\lambda^2\sigma^2}{V_{\textrm{min}}^{2/3}}-\sin^2\left(
\f{\lambda\sigma}{V_{\textrm{min}}^{1/3}}\right)\right]\approx 2V_{\textrm{min}}^{2}\,\rho_{\textrm{crit}}
\ee
Therefore, the volume at the bounce also approaches the $k$=0 value
\be
V_{\textrm{min}}=\frac{1}{\sqrt{2\rho_{\textrm{crit}}}}\,p_\phi
\label{V_min1f}
\ee
from above.

\begin{figure}[htb]
%\centerline{\includegraphics [height=5cm,width=6cm]{Fig4.pdf}}
\centerline{\includegraphics [scale=0.66]{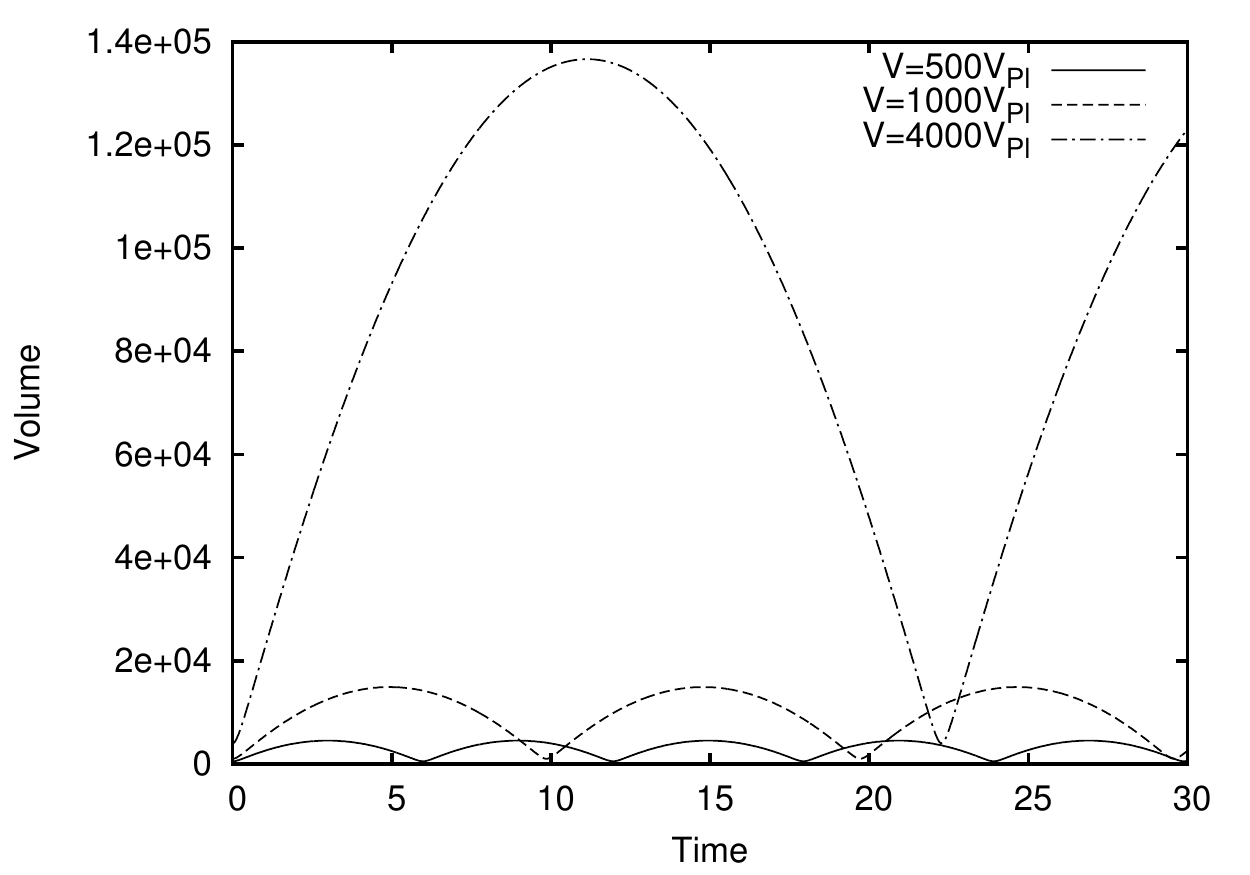}
\includegraphics [scale=0.66]{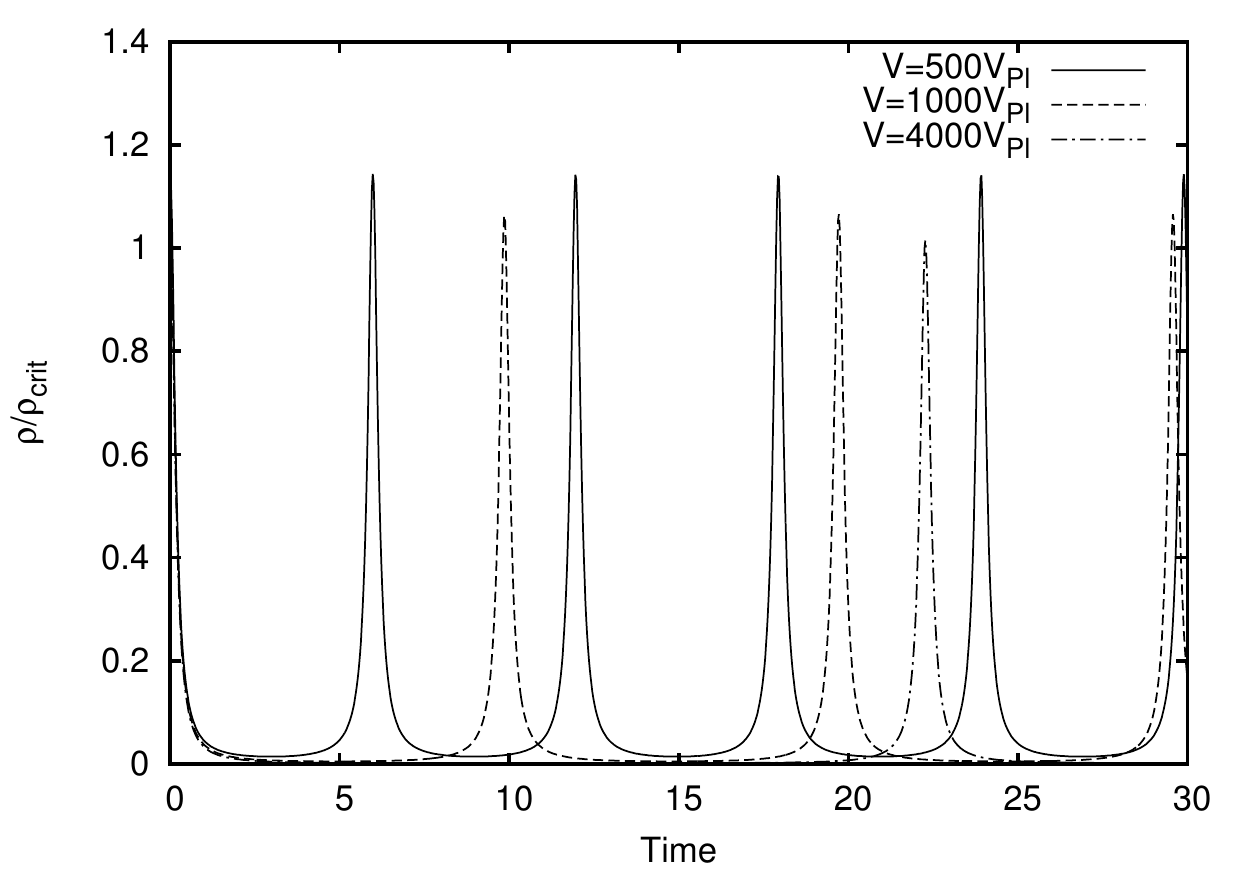}}
\caption{For three values of the volume at the bounce $V_{\textrm{b}}$, we plot the time evolution of the volume $V$ (left) and the density $\rho$ (right). These correspond to the values $V_{\textrm{b}}=500 \ell_{\textrm{Pl}}$ (---- line), $V_{\textrm{b}}=1000 \ell_{\textrm{Pl}}$ ($- - - - - $ line), and $V_{\textrm{b}}=4000 \ell_{\textrm{Pl}}$ ($- \cdot - \cdot -\cdot -$ line).}
\label{Fig:1}
\end{figure}

In Fig.~\ref{Fig:1} we have plotted the time evolution of three universes for three different values of volume $V_{\textrm{b}}$ at the bounce. From our previous expressions we see that the higher the value of the volume at the bounce, the higher the field momentum $p_\phi$ and the more semiclassical the trajectory. Note that this can be seen from the fact that the universe grows to larger values as one increases $p_\phi$, and the density at the bounce
{\it decreases} and tends to the value $\rho_{\textrm{crit}}$.

To summarize this section, we have seen that the effective dynamics of the holonomy based quantization, as defined in \cite{apsv}, yields a cyclic universe with a bounce at a matter densities that are larger than in the flat $k$=0 case. In the `large volume regime', the volume at which the expansion of the universe stops approaches the value given by general relativity. Through-out the evolution, a key geometrical scalar such as the expansion of cosmological observers remains absolutely bounded, and is saturated by all trajectories at the end of the superinflation regime that follows the bounce. These results complement those of  \cite{ps:fv} where it was shown that, within this quantization, singularity resolution in generic for a large class of matter.

%Note that for validity of equation for $V_{min}$, the density of matter field at the bounce has to be chosen in such a way that the minimum volume is in semi classical level (greater than Plank's volume).
%-----------------------------------------------------------------------

%%%%%%%%%%%%%%%%%%%%%%%%%%%%%%%%%%%%%%%%%%%%%%%%%%%%%%%%%%%%%%%%%%%%%%%%%%

\section{Loop quantization II: The connection way}
\label{sec:4}

For Bianchi II and IX cosmological models, where the spatial geometry has non trivial curvature, it was realized that the standard method of loop quantization based on
holonomies for closed loops, was not implementable in the Hilbert space of loop quantum cosmology. A new quantization prescription was put forward in \cite{bianchiII} and also employed in \cite{bianchiIX}. The basic idea is to define
an operator for the connection, by means of open holonomies, from which one can define the curvature. In this section we shall employ this quantization procedure to the closed $k$=1 FRW model.

To be precise, we  define the connection by an open holonomy, from which we arrive at the
expression for the connection
\be
A^i_a=\frac{\sin\bar\mu c}{\mu}\ ^o\omega^i_a
\ee
where $\bar\mu$ is the length of the curve which we use to calculate the holonomy along it and here we take $\bar\mu=\sqrt{\lambda^2/|p|}$. Just as the previous section, we shall use the variables $\beta$ and $V$ instead of $c$ and $p$.

This section has three parts. In the first one we derive the loop quantization for this prescription, writing in detail the quantum equations that define the theory when the matter is given by a massless scalar field. This
resulting formalism can then be directly compared to that of \cite{apsv}. In the second part, we consider the effective Hamiltonian and equations of motion derived from the quantum theory and analyze some of their general properties. In the last part we specialize in the massless scalar case where we can find explicit formulae for some of the relevant parameters of the solutions.

\subsection{Quantum Kinematics}

Let us start by recalling the classical Hamiltonian constraint,
\be
\H_{class}=-\frac{3}{8\pi G\gamma^2}\left[V\beta^2-2V^{2/3}\sigma\beta+V^{1/3}(1+\gamma^2)\sigma^2\right]+\rho V
\ee
where $\sigma=\ell_o/a_o=(2\pi)^{1/3}$ and $\rho=p_{\phi}^2/2V^2+U(\phi)$

As is standard in loop quantum cosmology, the gravitational part of the kinematical Hilbert space where the
constraints are to be implemented, is given by the so called {\it polymer Hilbert space} \cite{CVZ-2}. In that
Hilbert space, we can choose a basis of eigenstates,
\be
\hat v|v\rangle=v|v\rangle
\ee
which is related to the volume $\hat{V}$ as follows: $\hat{V}=\left(\f{8\phi\gamma}{6}\right)^{3/2}\f{|v|}{K}$ with
$K=2\sqrt{2}/(3\sqrt{3\sqrt{3}})$.
In this basis, $\exp i\lambda\beta$ becomes a translation operator.
\be
e^{i\lambda\beta/2}|v\rangle=|v+1\rangle
\ee
then
\be
\sin\lambda\beta|v\rangle=\frac{1}{2i}(|v+2\rangle-|v-2\rangle)
\ee
The quantum gravitational part of the Hamiltonian constraint operator is:
\be
\hat \H_{grav}=-\frac{3}{8\pi G\gamma^2\lambda^2}\left[\hat V^{1/4}\sin\lambda\beta\hat V^{1/2}\sin\lambda\beta\hat V^{1/4}-2\lambda\sigma\hat V^{1/3}\sin\lambda\beta\hat V^{1/3}+\lambda^2\sigma^2(1+\gamma^2)\hat V^{1/3}\right]
\ee
When the matter is given by a massless scalar field the quantum Hamiltonian constraint is
\be
\begin{split}
\hat \H=&-\frac{3}{8\pi G\gamma^2\lambda^2}\left[\hat V^{1/4}\sin\lambda\beta\hat V^{1/2}\sin\lambda\beta\hat V^{1/4}-2\lambda\sigma\hat V^{1/3}\sin\lambda\beta\hat V^{1/3}+\lambda^2\sigma^2(1+\gamma^2)\hat V^{1/3}\right]\\
&+\frac{\hat p_{\phi}^2}{2}\hat V^{-1}
\end{split}
\ee
To define the operator $\hat V^{-1}$, we first need to define $\widehat{|p|^{-1/2}}$ by means of Thiemann's prescription and, since $\widehat{|p|^{-1/2}}$ is well defined, then we can take its cube to define $\hat V^{-1}$,
\be
\widehat{|p|^{-1/2}}\Psi(v)=\frac{3^{5/6}\lambda}{2}|v|^{1/3}\left||v+1|^{1/3}-|v-1|^{1/3}\right|\Psi(v)
\ee
and then
\be
\hat V^{-1}\Psi(v)=\frac{\sqrt 3}{\lambda^3}f(v)
\ee
where
\be
f(v)=\left(\frac{3}{2}\right)^3 |v|\left||v+1|^{1/3}-|v-1|^{1/3}\right|^3\, .
\ee
The action of the Hamiltonian constraint operator on a state is given by
\be
-\hbar^2\partial_{\phi}^2\Psi(v;\phi)=\hat\Theta\Psi(v;\phi)
\ee
where the operator $\hat\Theta$ is given by
\be
\begin{split}
\hat\Theta\Psi(v;\phi)&=-\frac{2\sqrt 3f(v)^{-1}}{\lambda^3}\hat C\Psi(v;\phi)\\
&=-\frac{\sqrt 3^{1/3}\lambda^2}{8\pi G\gamma^2}[\frac{\lambda^2}{3^{1/3}}|v(v+4)|^{1/4}\frac{\sqrt{|v+2|}}{4}\Psi(v+4;\phi)-i\frac{\lambda^2\sigma}{3^{1/6}}|v(v+2)|^{1/3}\Psi(v+2;\phi)\\
&+[\frac{\lambda^2}{3^{1/3}}\frac{\sqrt{|v+2|}+\sqrt{|v-2|}}{4}-\lambda^2\sigma^2(1+\gamma^2)|v|^{1/3}]\Psi(v;\phi)\\
&-i\frac{\lambda^2\sigma}{3^{1/6}}|v(v-2)|^{1/3}\Psi(v-2;\phi)+\frac{\lambda^2}{3^{1/3}}|v(v-4)|^{1/4}\frac{\sqrt{|v-2|}}{4}\Psi(v-4;\phi)]
\end{split}
\ee
The final quantum theory has a structure very similar to that of \cite{apsv}. The non-separable Hilbert space
$\H_{\textrm{kin}}$ of the gravitational degrees of freedom is decomposed into an uncountable number, label by a parameter $\epsilon$, of superselected sectors $\H_\epsilon$, each of which is by itself, separable. The space of solutions can be given a Hilbert space
structure if one restricts attention to {\it positive frequency}, with respect to the internal time $\phi$. Thus
physical solutions $\psi$ satisfy the Schroedinger like equation,
\be
-i \partial_\phi\, \Psi = \sqrt{\hat\Theta}\,\Psi
\ee
A physical inner product can be defined on the space of solutions from which the physical Hilbert space can be constructed. An interesting avenue would be to perform a detailed analysis of the solutions of this theory, along the lines of \cite{apsv}. We shall leave that for future work. Let us now consider the effective description associated to the quantum theory described in this part.

\subsection{Effective Equations}

It is straightforward to see that the effective Hamiltonian one obtains from the quantum theory of the previous part, when neglecting inverse scale factor effects (as was done in \cite{apsv} and \cite{ps:fv}), is
\be
\mathcal{H}_{\textrm{eff}}=-\frac{3}{8\pi
G\gamma^2\lambda^2}V\left[(\sin\lambda\beta-D)^2+\gamma^2 D^2\right]+\rho V\, .
\ee
It is then straightforward to compute the corresponding effective equations of motion.
In particular, by computing $\dot{V}=\{V,\mathcal{H}_{\textrm{eff}}\}$, we can find the expression
for the expansion as
\be
\theta=\frac{3}{\lambda\gamma}\cos\lambda\beta\left(\sin\lambda\beta-D\right)\label{exp-2}\, .
\ee
From which we can find the effective Friedman equation,
\be
H^2=\frac{1}{\lambda^2\gamma^2}\cos^2\lambda\beta\left(\sin\lambda\beta-D\right)^2=\frac{8\pi G}{3}(\rho-\rho_1)(1-\frac{\rho-\rho_2}{\rho_{\textrm{crit}}})\label{frid-2}\, ,
\ee
where $\rho_1=\rho_{\textrm{crit}}\gamma^2D^2$  and $\rho_2=\rho_{crit}D[(1+\gamma^2)D-2\sin\lambda\beta]$.
Let us now explore what is the difference in the behavior of the Universe as described by these equations,
compared to the dynamics given by the holonomy-based quantization. The first obvious observation from Eq.~(\ref{exp-2}) is that the universe undergoes a turnaround whenever the expansion vanishes. This can happen
either when: a) $\sin\lambda\beta= D$, or b) when $\cos\lambda\beta=0$. The first condition can also be written, by using (\ref{frid-2}), as $\rho=\rho_1=\rho_{\textrm{crit}}\gamma^2D^2$, and in the
limit $D\ll 1$ --when the volume is large in Planck units-- corresponds to the point of re-collapse.
It is interesting to note that, in contrast to the other quantum theory, the expression for the point of re-collapse here coincides {\it exactly} with that of the classical theory (recall that in the previous case, we only recovered this value in the large volume/momentum limit).

Just as we had in the previous case, we expect that the nature of the turnaround points (whether they correspond to a bounce or a re-collapse) will be determined only after we consider the rate of change of the expansion (the Hubble).
The second condition above, namely condition b) can be written as $\rho = \rho_{crit} + \rho_2$, or alternatively, as
$\cos\lambda\beta=0$.
Now, for this condition ``b)'', there is a crucial difference with the previous case. While in the effective description of the holonomy based quantization all equations were invariant under the mapping $\beta \to \beta + \pi/\lambda$ (and therefore
implementing an effective periodicity of $\beta$ with period $\pi/\lambda$), this is no longer the case here.
Even when the zeros of the term $\cos\lambda\beta$ have that periodicity, the term $\sin\lambda\beta - D$ does not. Therefore,
there are two kind of roots for the equation $\cos\lambda\beta=0$. The first root  `b.1' occurs when $\beta_n=\f{(4n+1)\pi}{2\lambda}$, where $\sin\lambda\beta_n=1$. The other root `b.2' is when $\beta_m=\f{(4m+3)\pi}{2\lambda}$, in which case $\sin\lambda\beta_m= -1$. The important thing here to
notice is that the density (and therefore, volume) are different in these two cases, which implies that {\it there
are two different kind of turnarounds of type `b)'}.

In order to identify the nature of these turnaround point, let us use the rest of the equations of motion,
%Let us now write the rest of the equations of motion.
\be
\dot\beta =4\pi G\gamma P-
\frac{1}{2\gamma\lambda^2}\left[3\sin^2\lambda\beta-4D\sin\lambda\beta+(1+\gamma^2)D^2\right]\, ,
%&=-\frac{4\pi G\gamma p_\phi^2}{V^2}+\frac{1}{\gamma\lambda^2}\left[(1+\gamma^2)D^2-D\sin\lambda\beta\right]
%\end{split}
\ee
and, from the continuity equation, we get
\be
\dot\beta=-4\pi G\gamma(\rho-\rho_3+P)\quad \textrm{where}\quad\rho_3=\frac{2\rho_{\textrm{crit}}D}{3}\left[(1+\gamma^2)D-\sin\lambda\beta\right]
\ee
Finally, we have the change of the expansion function given as
\be
\dot\theta=\frac{3}{\gamma}\dot\beta\left(\cos2\lambda\beta+D\sin\lambda\beta\right)+\frac{D\theta}{\lambda\gamma}\cos\lambda\beta\label{dot-theta2}
\ee
From this last equation we can then determine the identity of the turnaround points.
 For the different cases as defined  above we have,

\noindent
Case a): It is defined by $\sin\lambda\beta=D$, or alternatively by $\rho=\tilde\rho_1=\rho_{\textrm{crit}}\gamma^2D^2$.
In this case,
\be
\dot\theta=\f{3}{\gamma}\,\dot\beta (\cos^2\lambda\beta-\sin^2\lambda\beta+D\sin\lambda\beta)
=\f{3}{\gamma}\,\dot\beta\,\cos^2\lambda\beta
\ee
Thus, just as it happened in the holonomy-based quantization, when $\dot\beta < 0$ this point corresponds to
a re-collapse, while in the case that $\dot\beta > 0$, this is a bounce.

\noindent
Case b): It is defined by $\cos\lambda\beta=0$, or equivalently by $\rho=\rho_{\textrm{crit}}[1 + D((1 + \gamma^2)D-2\sin\lambda\beta)]$. In this case we have two subcases, corresponding to the two roots of the
equation $\cos\lambda\beta=0$.\\
Case b.1) This corresponds to the roots $\lambda\beta_n=\frac{(4n+1)\pi}{2\lambda}$, for $n$ integer. In this case,
$\sin\l\beta_n=1$, so the change of the expansion in given by,
\be
\dot\theta_1=-\f{3}{\gamma}\,\dot\beta\, (1 - D)
\ee
Thus, we see that the nature of the turnaround depends not only on the sign of $\dot\beta$ but also on the
magnitude of $D$. In the large volume regime, where $D\ll 1$, we have the same situation as in the holonomy-based quantization, namely that in the $\dot\beta<0$ case, the turnaround point corresponds to a {\it bounce} (and
in the $\dot\beta > 0$ case, to a {\it re-collapse}). The density is given then by,
\be
\rho_{\textrm b}^{1}=\rho_{\textrm{crit}}\left[(1-D)^2+\gamma^2D^2\right]\, ,
\ee
Let us nos consider the other root.\\
Case b.2) This corresponds to the root $\lambda\beta_m=\frac{(4m+3)\pi}{2}$ for $m$ integer. In this case,
$\sin\l\beta_n= - 1$, so the change of the expansion in given by,
\be
\dot\theta_2=-\f{3}{\gamma}\,\dot\beta\, (1 + D)\, .
\ee
We have the same situation as in the holonomy-based quantization, namely that in the $\dot\beta<0$ case, the turnaround point corresponds to a {\it bounce} (and
in the $\dot\beta > 0$ case, to a {\it re-collapse}). The density is given then by,
\be
\rho_{\textrm b}^{2}=\rho_{\textrm{crit}}\left[(1+D)^2+\gamma^2D^2\right]\, .
\ee

To summarize, instead of two turnaround points as in the holonomy-based quantization, this new quantization has the
novel feature that there are three different turnaround points. In the case of large volume and for $\dot\beta < 0$,
they correspond to {\it two bounces} and a re-collapse. For extreme situations near the Planck scale and for certain matter content one might have different scenarios \cite{CKM}.

%  the solution of bellow equation is the maximum volume for this case
%\be
%\frac{p_\phi^2}{2V_{max}^2}+U(\phi)=\rho_{crit}\frac{\gamma^2\lambda^2\sigma^2}{V^{2/3}}
%\label{V_max20}
%\ee

%From above equations we can see when $\rho = \rho_{crit} + \rho_2$ (or equivalently $\lambda\beta = (2n + 1)\pi/2$)
%bounce occurs and we have recollapse when $\rho =\rho_1=\rho_{crit}\gamma^2D^2$ (or equivalently $\sin\lambda\beta = D$).

\subsection{An example: A massless scalar}

Let us now consider
as matter field a massless scalar field $\phi$, for which $\dot\beta<0$ and does not change sign during the
dynamical evolution. Furthermore, we shall assume $D < 1$, in which case, the case a) above corresponds to the
point of re-collapse, while the points b.1) and b.2) correspond to the two distinct {\it bounces}.
The maximum value of volume is exactly given by,
\be
V_{\textrm{max}}=\left(\frac{64\pi G}{3\sigma^2}\right)^{3/4}p_\phi^{3/2}
\label{V_max2f}
\ee
which is equal to the classical value for maximum volume for the FRW model with $k$=1.
The equations for minimum volumes which correspond to the two different bounces are
\be
\frac{p_{\phi}^2}{2V_{\textrm{min}}^2} = \rho_{\textrm{crit}}\left[(1+\frac{\lambda\sigma}{V_{\textrm{min}}^{1/3}})^2+\frac{\gamma^2\lambda^2\sigma^2}{V^{2/3}_{\textrm{min}}}\right]
\label{V_min210}
\ee
and
\be
\frac{p_{\phi}^2}{2V_{\textrm{min}}^2} =
\rho_{\textrm{crit}}\left[(1-\frac{\lambda\sigma}{V_{\textrm{min}}^{1/3}})^2+\frac{\gamma^2\lambda^2\sigma^2}{V^{2/3}_{min}}\right]
\label{V_min220}
\ee
%-------------------------------------------------------------------------
\par
In the limit of large field's momentum $p_\phi$, since the volume is also large then we have $D\ll 1$.
We can write the density at the two bounces as follows,
$$\rho_{\textrm{b}}^{1}=\rho_{\textrm{crit}}\left[(1+D)^2+\gamma^2D^2\right]\quad{\textrm{and}}\quad
\rho_{\textrm{b}}^{2}=\rho_{\textrm{crit}}\left[(1-D)^2+\gamma^2D^2\right]\, ,$$
from which it follows that, in the limit $D\ll 1$ they both tend to $\rho_{\textrm{crit}}$ from above.
Therefore the density at the bounce for both approaches with different quantization in this limit approaches $\rho_{\textrm{crit}}$ the critical density for the $k$=0 FRW model.
Since both bounce densities have the same limit, then the minimum value of the volume for both cases goes to
\be
V_{\textrm{min}}\approx \sqrt\frac{1}{2\rho_{\textrm{crit}}}\;p_\phi
\label{V_min2f}
\ee
therefore, when the field's momentum $p_\phi$ is very large, since we can ignore the negative powers of volume, the maximum absolute value of expansion for the second approach goes to $3/2\gamma$ which is the same as in first approach.

\begin{figure}[htb]
%\centerline{\includegraphics [height=5cm,width=6cm]{Fig4.pdf}}
\centerline{\includegraphics [scale=0.66]{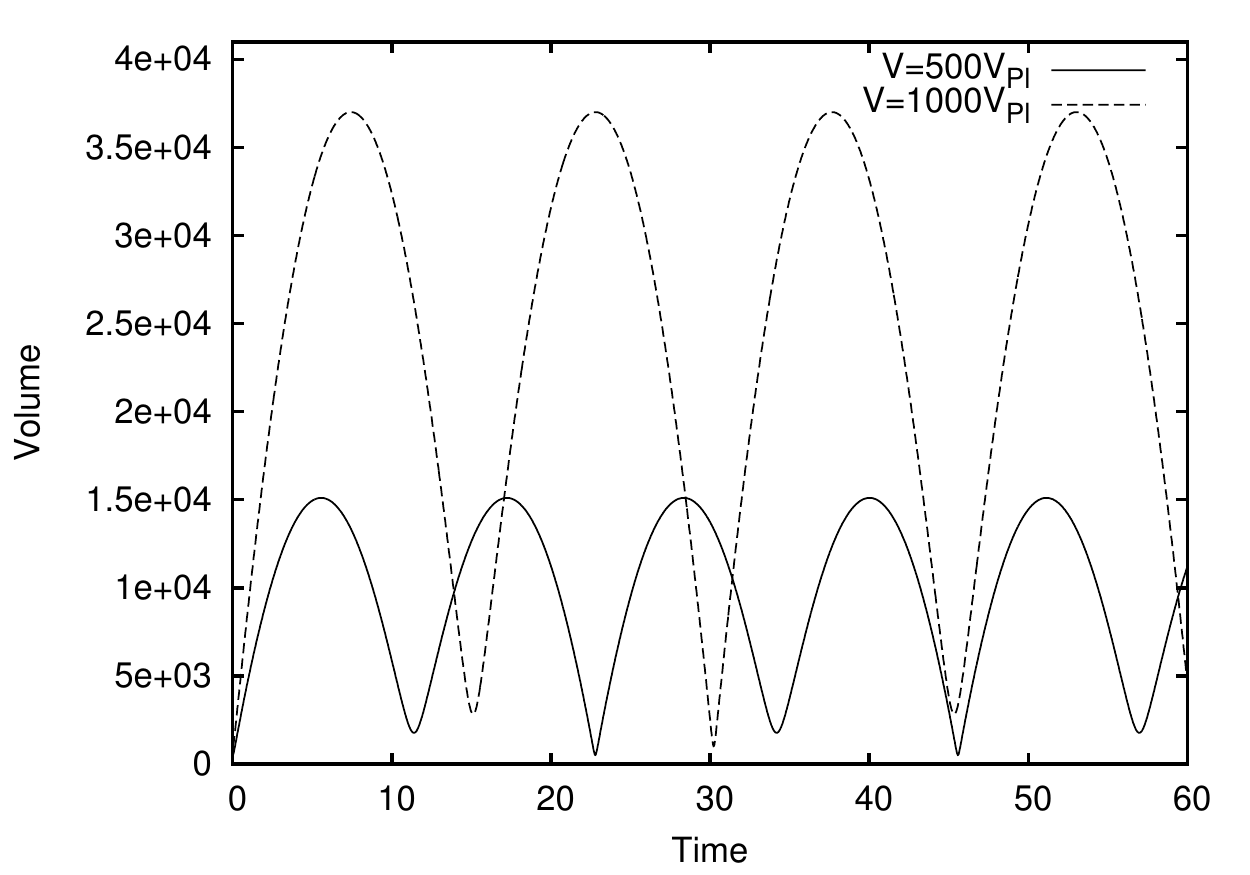}
\includegraphics [scale=0.66]{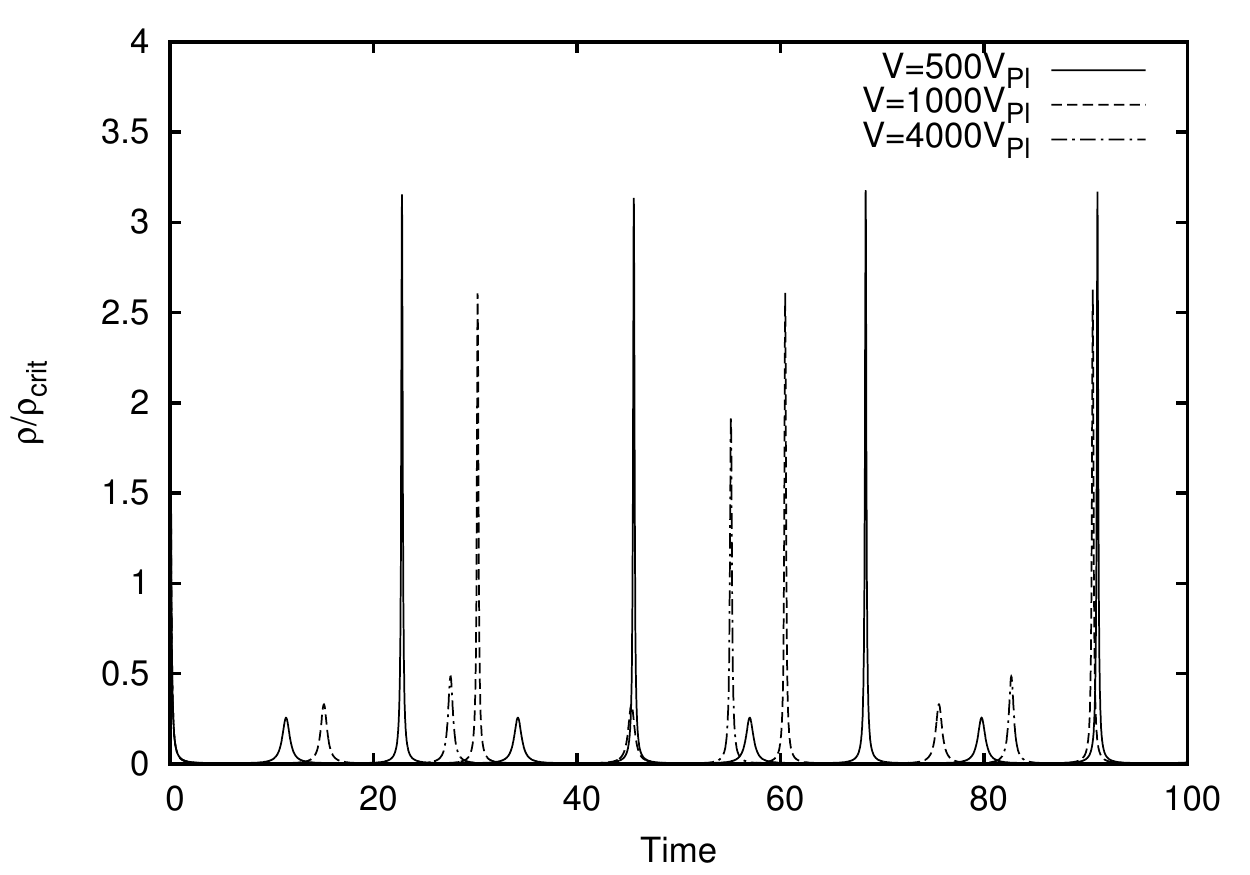}}
\caption{For three values of the volume at the bounce $V_{\textrm{b}}$, we plot the time evolution of the volume $V$ (left) and the density $\rho$ (right). These correspond to the values $V_{\textrm{b}}=500 \ell_{\textrm{Pl}}$ (---- line), $V_{\textrm{b}}=1000 \ell_{\textrm{Pl}}$ ($- - - - - $ line), and $V_{\textrm{b}}=4000 \ell_{\textrm{Pl}}$ ($- \cdot - \cdot -\cdot -$ line).}
\label{Fig:2}
\end{figure}

\begin{figure}[htb]
%\centerline{\includegraphics [height=5cm,width=6cm]{Fig4.pdf}}
\centerline{\includegraphics [scale=0.66]{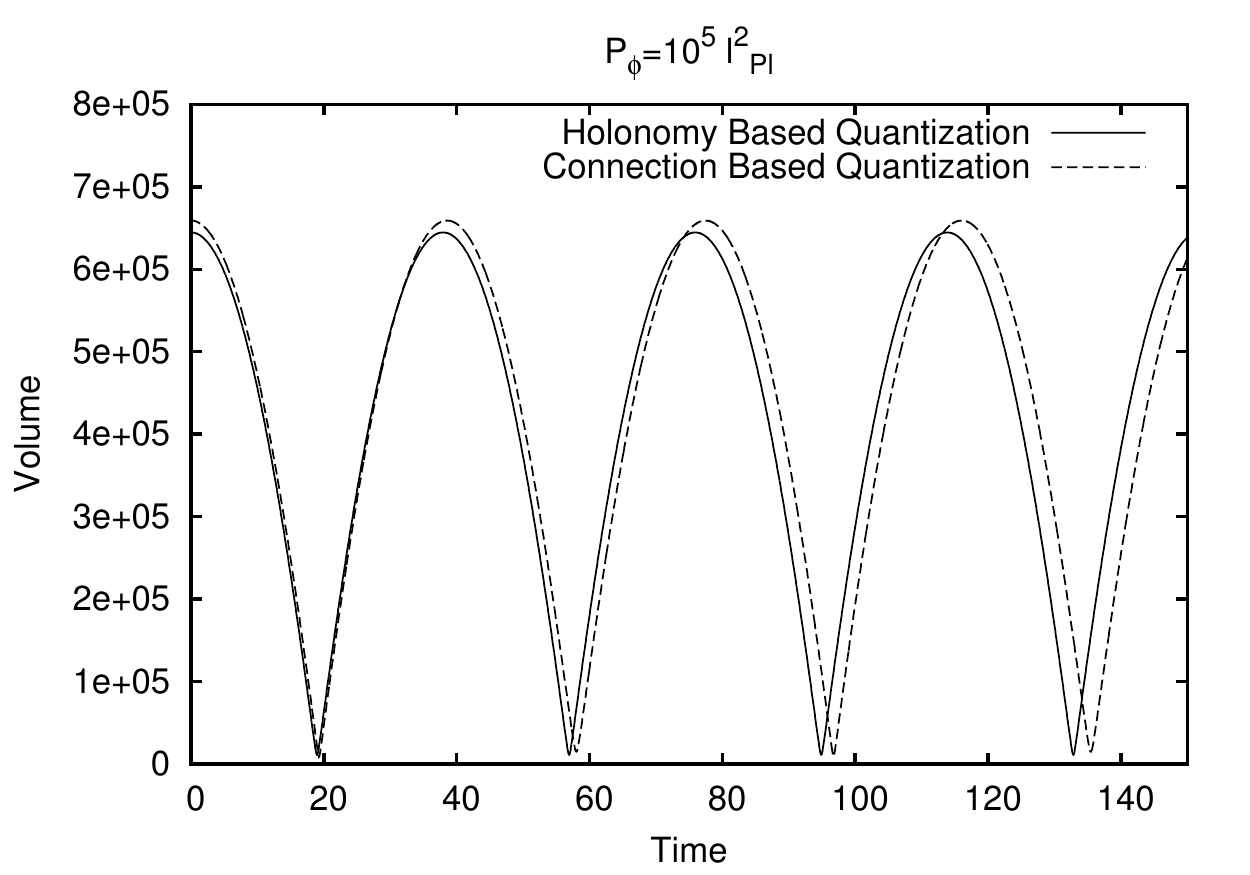}
\includegraphics [scale=0.66]{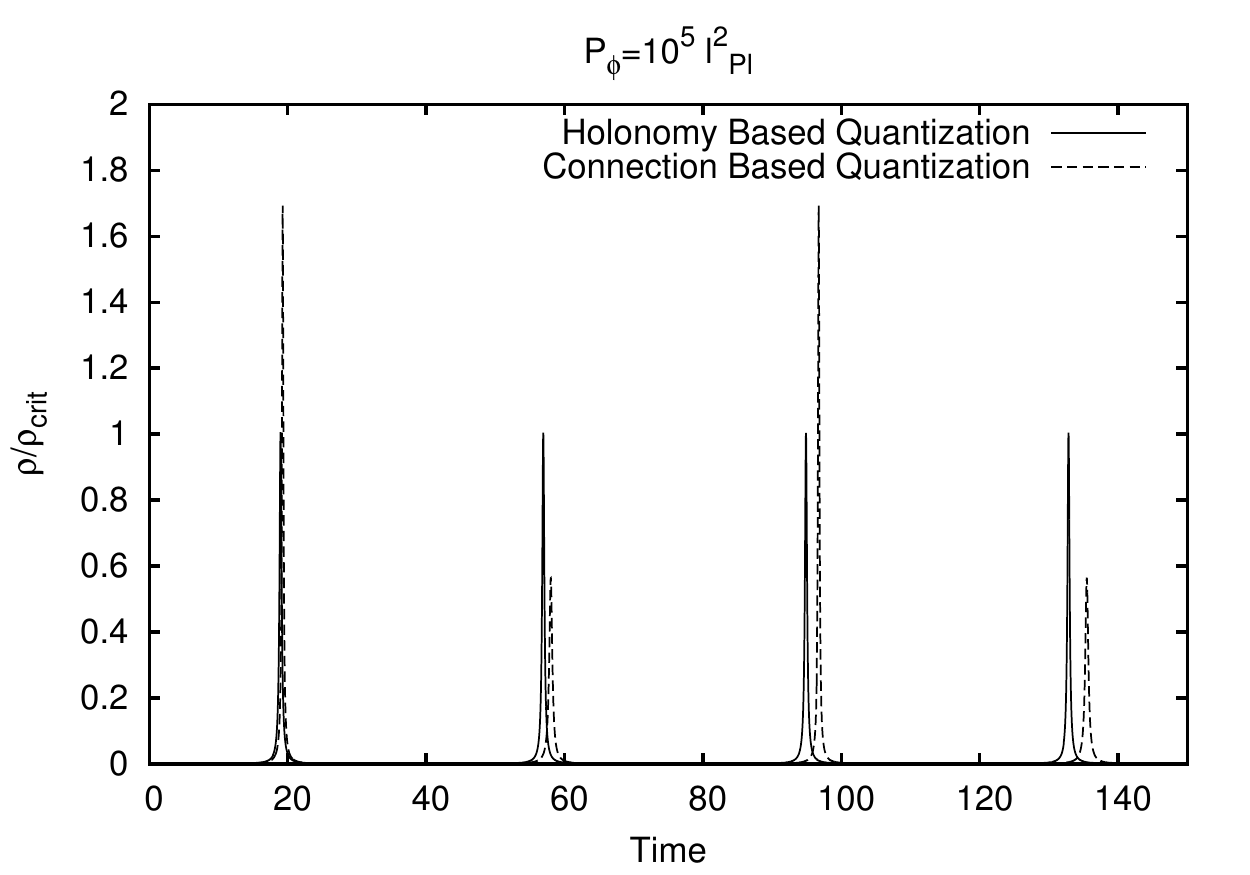}}
\caption{We plot the time evolution of the volume $V$ (left) and the density $\rho$ (right), for the two
quantization methods, for $p_\phi=10^5$.}
\label{Fig:3}
\end{figure}

In Fig.~(\ref{Fig:2}) we have plotted the time evolution of the universe for different values of the minimum volume at the bounce. As we can see, as we increase this value, and therefore, the field's momentum $p_\phi$, the two bounces tend to each other, both in terms of the value of the volume and in the maximum value of the densities. Note  that the densities at the `strongest' bounce are much higher, in this regime, than in the holonomy-based quantization, and that they decrease as one increases the value of $p_\phi$.
One can further compare both description by fixing the value of $p_\phi$ and comparing the time evolution of
volume and density. We have plotted such comparison in Fig.~(\ref{Fig:3}) for $p_\phi=10^5$. Note that the density at the bounce in the holonomy-based quantization is in between the two densities for the connection-based quantization. The period between the point of re-collapse is not the same for both schemes but, as one increases $p_\phi$,
they approach each other, just at the volume and density at the bounce converge.

Let us summarize the results this section. First, we developed the quantum theory for $k$=1 loop quantum gravity coupled to a scalar field, employing a quantization method the uses open holonomies to regulate the field
strength appearing in the constraint. In the second part we derived some of the consequences of such a quantum theory, by means of its effective description. We found that the most dramatic difference from the quantization
previous explored is that the cyclic universe undergoes cycles of contraction and expansion, but alternating between two different quantum bounces (or alternating between two kinds of points of re-collapse and a bounce). Furthermore, we saw that for `large universes', where the universe expands to a large volume (in Planck units), the densities (and volumes) of the two distinct bounces approach each other and converge to
the values attained in the $k$=0 theory.
%-------------------------------------------------------------------------

\section{Discussion}
\label{sec:5}

In this article we have explored a quantization ambiguity that exists for certain models in loop quantum cosmology.
This correspond to the freedom of using closed holonomies around loops to define curvature or open holonomies to define connections. Since it is only the latter choice that is available for anisotropic models with non-trivial spatial curvature, it is important to understand the particular features of this quantization, and compare it to the
original holonomy-based loop quantization. In this regard, the isotropic $k$=1 FRW model is ideal since both quantizations exist and are not equivalent (while they are in the case of $k$=0 and Bianchi I). We have explored some of the differences between these two theories, by means of their corresponding {\it effective} descriptions. The equation of motion for both theories are not the same, and therefore their underlying dynamics is different. The most dramatic difference is that, while
the universe is cyclic in the holonomy-based quantization with a bounce followed by a re-collapse, in the new
quantization the situation is more complicated, with three different turnaround points. In the semiclassical limit
where the universe is a assumed to grow large, we have seen that there
 are {\it two} kinds of bounces with different densities that alternate with the re-collapse.
The volume at which the expansion stops and the universe starts to contract is also different.

Interestingly, in the limit of large universes both theories converge and the two distinct bounces
of the connection-based theory approach that of the
holonomy-based quantization. In this limit both descriptions approximate general relativity during the small density epochs of the cyclic universes, making them almost indistinguishable.
It would be interesting to explore further the similarities and differences of the two approaches regarding singularity resolution, as was done in \cite{ps:fv} for the holonomy based description. Further numerical analysis with various matter fields might yield significant differences that could have potential observable consequences. This shall be reported elsewhere \cite{CKM}.

\section*{Acknowledgments}

\noindent We would like to thank P. Singh for
discussions and comments, and E. Montoya for discussions and for help with the figures. This work was in part supported by DGAPA-UNAM IN103610 grant, by NSF
PHY0854743 grant and by the Eberly Research Funds of Penn State.

\begin{appendix}

\section{The three sphere, holonomies and curvature}

For a 3-sphere with radius equal to $a_o$, the line element can be written as $$\dif s^2=a_o^2(\dif\alpha^{\prime 2}+\dif\beta^{\prime 2}+\dif\gamma^{\prime 2}+2\cos\beta\dif\alpha^\prime\dif\gamma^\prime)$$
 where $0\leq\alpha^\prime\leq\pi$, $0\leq\beta^\prime\leq\pi/2$ and $0\leq\gamma^\prime\leq 2\pi$.
 With a simple redefinition of coordinates,   $\alpha=2\alpha^\prime$, $\beta=2\beta^\prime$ and $\gamma=2\gamma^\prime$, it can be written as
\be
\dif s^2=\frac{a_o^2}{4}(\dif\alpha^2+\dif\beta^2+\dif\gamma^2+2\cos\beta\dif\alpha\dif\gamma)\label{fid-met}
\ee
where $0\leq\alpha\leq 2\pi$, $0\leq\beta\leq\pi$ and $0\leq\gamma\leq 4\pi$. For this metric, the
volume of $\Sigma$ is $V_0=2\phi^2\,a_0$.  Recall that we have defined $\ell_o=V_o^{1/3}$, and $\sigma=\ell_o/a_o=(2\pi^2)^{1/3}$.
\par

Let us now compute the holonomy along the edge $e$ with length $\ell^\prime$, parameterized by $\ell$, tangential to vector $t^a=(\partial/\partial \ell)^a$. It is given by
\be
h^{(\mu)}=\exp(\int_e A \cdot\textrm{d}e(\ell))=\exp(\int^{\ell^\prime}_{0}t^aA^{j}_{a}\tau_{j}\d\ell)\, .
\ee
If we want to use some angular parameters like $\theta$ instead of $\ell$ we will have, for a general integral,
\be
\int_0^{\ell^\prime} \textrm{d}\ell\textrm{ } t(F)=\int_0^{\ell^\prime/a}\textrm{d}\theta \textrm{ }t^\prime(F)
\ee
with $t^\prime=\frac{\partial}{\partial\theta}$ and $a$ playing the role of a `radius', since $\ell=a\,\theta$.
For our problem, we can define
$$
t^\prime=\pm\frac{a_o}{2}\,{}^o\!e_3=\pm\frac{\ell_{o}}{2\sigma}\,{}^o\!e_3\quad\textrm{or}\quad\pm\frac
{a_o}{2}\xi_3=\pm\frac{\ell_{o}}{2\sigma}\, \xi_3\, .
$$
Therefore, to calculate a component of $F_{ab}^k$ of the curvature, we can construct a closed loop as follows. In  coordinates $(\alpha,\beta,\gamma)$\\
i) Move from $(0,\pi/2,0)$ to $(0,\pi/2,2\sigma\mu)$ following ${}^oe_3=\partial/\partial\gamma$,\\
ii) Then move from $(0,\pi/2,2\sigma\mu)$ to $(2\sigma\mu,\pi/2,2\sigma\mu)$ following $-\xi_3=\partial/\partial\alpha$,\\
iii) Next, move from $(2\sigma\mu,\pi/2,2\sigma\mu)$ to $(2\sigma\mu,\pi/2,0)$ following $-{}^oe_3$, and finally\\ iv) Move from $(2\sigma\mu,\pi/2,0)$ to $(0,\pi/2,0)$ following $\xi_3$.

The open holonomy along one edge, with parameter $\mu$ is given by
\be
h^{(\mu)}=\exp(\int^{2\mu\ell_o/a_o}_0 t^{\prime a}A^{j}_{a}\tau_{j}\d\theta)
\ee
where $\theta=\alpha \textrm{ or }\gamma$ depending on the edge, and the effective radius of the 3-sphere used to translate from lengths to angles is $a_0/2$ (compatible with the fiducial metric (\ref{fid-met})). Thus, we will have for the closed loop defined above,
\be
\begin{split}
h_{\Box_{31}}&=h_4h_3h_2h_1=e^{\tau_1\mu c}e^{-\tau_3\mu c}e^{-(\sin(2\sigma\mu)\tau_2+\cos(2\sigma\mu)\tau_1)\mu c}e^{\tau_3\mu c}
\end{split}
\ee
then we have
\be
{}^oe_3^a{}^oe_1^bF^k_{ab}=\lim_{\mu\rightarrow 0}\frac{2}{\mu^2\ell_o^2}\textrm{Tr}(h_{\Box_{31}}\tau^k)=-\frac{1}{\ell_o^2}(c^2-2\sigma c)
\ee
recovering thus the classical expression for curvature. If we do not take the limit $\mu\to 0$ but instead
take the area as the smallest eigenvalue of the area operator, or equivalently $\bar\mu^2|p|=\lambda^2$ then
the curvature can be approximated, at scale $\lambda$, as
\be
{}^{\lambda}\!F_{ab}^k=\frac{\sin^2\bar\mu(c-\sigma)-\sin^2(\bar\mu\sigma)}{\bar\mu^2\ell_o^2}
\ee
where $\bar\mu=\sqrt{\lambda^2/|p|}$.
%we can change the variables from $c$ and $p$ to $\beta=c|p|^{-1/2}$ and $V=|p|^{3/2}$ (the volume).

\end{appendix}


\begin{thebibliography}{99}

\bibitem{lqc} M.~Bojowald,
 ``Loop quantum cosmology",
 \emph{Living Rev.\ Rel.}\  {\bf 8}, 11 (2005)
 {\tt arXiv:gr-qc/0601085}.
 %%CITATION = GR-QC 0601085;%%

\bibitem{abl} A.~Ashtekar, M.~Bojowald and L.~Lewandowski,
 ``Mathematical structure of loop quantum cosmology"
 \emph{Adv.\ Theor.\ Math.\ Phys.}\  {\bf 7} 233 (2003)
 {\tt arXiv:gr-qc/0304074}.
 %%CITATION = GR-QC 0304074;%%

\bibitem{AA} A.~Ashtekar,
  ``Loop Quantum Cosmology: An Overview,''
  Gen.\ Rel.\ Grav.\  {\bf 41}, 707 (2009)
  {\tt arXiv:0812.0177 [gr-qc]}.
  %%CITATION = GRGVA,41,707;%%

\bibitem{lqg} A.~Ashtekar and J.~Lewandowski
``Background independent quantum gravity: A status report,''
Class.\ Quant.\ Grav.\  {\bf 21}  (2004)  R53
 {\tt arXiv:gr-qc/0404018};
%%CITATION = GR-QC 0404018;%%
C.~Rovelli,  ``Quantum Gravity", (Cambridge U. Press, 2004);
T.~Thiemann, ``Modern canonical quantum general
relativity,'' (Cambridge U. Press, 2007).
%%CITATION = GR-QC 0110034;%%


\bibitem{aps0} A.~Ashtekar, T.~Pawlowski and P.~Singh,
``Quantum Nature of the Big Bang,'' Phys. Rev. Lett {\bf 96} (2006) 141301 {\tt arXiv:gr-qc/0602086}.

\bibitem{aps1} A.~Ashtekar, T.~Pawlowski and P.~Singh,
``Quantum Nature of the Big Bang: An Analytical and Numerical Investigation,''
Phys. Rev. D {\bf 73} (2006) 124038. {\tt arXiv:gr-qc/0604013}.

\bibitem{aps2} A.~Ashtekar, T.~Pawlowski and P.~Singh,
 ``Quantum nature of the big bang: Improved dynamics,''
 Phys.\ Rev.\ D {\bf 74}, 084003 (2006)
 {\tt arXiv:gr-qc/0607039}.
 %%CITATION = GR-QC 0607039;%%





\bibitem{slqc} A.~Ashtekar, A.~Corichi and P.~Singh,
``Robustness of key features of loop quantum cosmology,''
Phys.\ Rev.\ D {\bf 77}, 024046 (2008). {\tt arXiv:0710.3565 [gr-qc]}.
%%CITATION = PHRVA,D77,024046;%%

\bibitem{cs:prl} A.~Corichi and P.~Singh,
  ``Quantum bounce and cosmic recall,''
  Phys.\ Rev.\ Lett.\  {\bf 100}, 161302 (2008)
  {\tt arXiv:0710.4543 [gr-qc]};
  %%CITATION = PRLTA,100,161302;%%


\bibitem{polish} W.~Kaminski and J.~Lewandowski,
 The flat FRW model in LQC: the self-adjointness,
 Class.\ Quant.\ Grav.\  {\bf 25}, 035001 (2008);
 {\tt arXiv:0709.3120 [gr-qc]}.

\bibitem{recall}  W.~Kaminski and T.~Pawlowski,
  ``Cosmic recall and the scattering picture of Loop Quantum Cosmology,''
  Phys.\ Rev.\  D {\bf 81}, 084027 (2010)
  {\tt arXiv:1001.2663 [gr-qc]}.
  %%CITATION = PHRVA,D81,084027;%%

\bibitem{CM} A.~Corichi and E.~Montoya,
  ``On the Semiclassical Limit of Loop Quantum Cosmology,''
  arXiv:1105.2804 [gr-qc];
  %%CITATION = ARXIV:1105.2804;%%
``Coherent semiclassical states for loop quantum cosmology" (preprint).


\bibitem{cs:unique} A.~Corichi and P.~Singh,
  ``Is loop quantization in cosmology unique?,''
  Phys.\ Rev.\  D {\bf 78}, 024034 (2008)
  {\tt arXiv:0805.0136 [gr-qc]};
  %%CITATION = PHRVA,D78,024034;%%
  ``A geometric perspective on singularity resolution and uniqueness in loop
  quantum cosmology,''
  Phys.\ Rev.\  D {\bf 80}, 044024 (2009)
  {\tt arXiv:0905.4949 [gr-qc]}.
  %%CITATION = PHRVA,D80,044024;%%


\bibitem{apsv} A. Ashtekar, T. Pawlowski, P. Singh, K. Vandersloot, ``Loop quantum cosmology of k=1 FRW models,''
Phys.Rev. D {\bf 75} (2007) 024035; {\tt arXiv:gr-qc/0612104}.


\bibitem{closed} L. Szulc, W. Kaminski, J. Lewandowski, ``Closed FRW model in Loop
Quantum Cosmology,''
Class.Quant.Grav. {\bf 24} (2007) 2621; {\tt arXiv:gr-qc/0612101}.

\bibitem{open} K. Vandersloot, ``Loop quantum cosmology and the k = - 1 RW model,''
Phys.Rev. D {\bf 75} (2007) 023523; {\tt arXiv:gr-qc/0612070}.

\bibitem{cosmos} P.~Singh,
  ``Are loop quantum cosmos never singular?,''
  Class.\ Quant.\ Grav.\  {\bf 26}, 125005 (2009)
  [arXiv:0901.2750 [gr-qc]].
  %%CITATION = CQGRD,26,125005;%%

\bibitem{ps:fv} P.~Singh and F.~Vidotto,
  ``Exotic singularities and spatially curved Loop Quantum Cosmology,''
  Phys.\ Rev.\  D {\bf 83}, 064027 (2011)
  [arXiv:1012.1307 [gr-qc]].
  %%CITATION = PHRVA,D83,064027;%%

\bibitem{vp}
E.~Bentivegna and T.~Pawlowski,
 ``Anti-deSitter universe dynamics in LQC,''
 {\tt arXiv:0803.4446 [gr-qc]}.
 %%CITATION = ARXIV:0803.4446;%%


\bibitem{massive} A. Ashtekar, T. Pawlowski, P. Singh, (In preparation).


\bibitem{bianchiI} A.~Ashtekar and E.~Wilson-Ewing,
  ``Loop quantum cosmology of Bianchi I models,''
  Phys.\ Rev.\  D {\bf 79}, 083535 (2009)
  [arXiv:0903.3397 [gr-qc]].
  %%CITATION = PHRVA,D79,083535;%%

\bibitem{bianchiII} A.~Ashtekar and E.~Wilson-Ewing,
  ``Loop quantum cosmology of Bianchi type II models,''
  Phys.\ Rev.\  D {\bf 80}, 123532 (2009)
  [arXiv:0910.1278 [gr-qc]].
  %%CITATION = PHRVA,D80,123532;%%

\bibitem{CVZ-2}
A.~Corichi, T.~Vukasinac and J.A.~Zapata, ``Polymer quantum
mechanics and its continuum limit", Phys Rev {\bf D76}, 044016
(2007). {\tt arXiv:0704.0007v1 (gr-qc)}.
%%CITATION = PHRVA,D76,044016;%%

\bibitem{bianchiIX}
E.~Wilson-Ewing,
  ``Loop quantum cosmology of Bianchi type IX models,''
  Phys.\ Rev.\  D {\bf 82}, 043508 (2010)
  [arXiv:1005.5565 [gr-qc]].
  %%CITATION = PHRVA,D82,043508;%%

\bibitem{ma}
  J.~Yang, Y.~Ding, Y.~Ma,
  ``Alternative quantization of the Hamiltonian in isotropic loop quantum cosmology,''
  [arXiv:0902.1913 [gr-qc]].



%\bibitem{victor} V. Taveras, ``Corrections to the Friedman equations from LQG for a Universe
%with a free scalar field",
%Phys.\ Rev.\  D {\bf 78}, 064072 (2008)
%  {\tt arXiv:0807.3325 [gr-qc]}.
  %%CITATION = PHRVA,D78,064072;%%



\bibitem{CKM} A.~Corichi, A.~Karami and E.~Montoya, to be published.



\end{thebibliography}
\end{document}